\documentclass[runningheads]{llncs}
\usepackage[T1]{fontenc}
\usepackage{graphicx}
\usepackage{hyperref}
\usepackage{color}

\usepackage{amsmath}
\usepackage{bm}
\newcommand{\sbigotimes}{%
  \mathop{\mathchoice{\textstyle\bigotimes}{\bigotimes}{\bigotimes}{\bigotimes}}%
}
\usepackage{braket}
\usepackage{float}
\usepackage{fontawesome5}

\begin{document}
\title{Efficient Classical Computation of Single-Qubit Marginal Measurement Probabilities to Simulate Certain Classes of Quantum Algorithms}
%
%
\author{
Santana Yuda Pradata\inst{1,2}\orcidID{0009-0009-0470-8456} \and
Muhammad Anin Nabail Azhiim\inst{2}\orcidID{0009-0002-1040-8571} \and
Hendry Minfui Lim\inst{2}\orcidID{0000-0002-2957-2438} \and
Wiwit Suryanto\inst{3}\orcidID{0000-0002-6275-7880} \and
Ahmad Ridwan Tresna Nugraha\inst{2}\orcidID{0000-0002-5108-1467} \and
Muhammad Alfian Amrizal\inst{1}\orcidID{0000-0003-1124-5137}\faEnvelope[regular] \and
Hiroyuki Takizawa\inst{4}\orcidID{0000-0003-2858-3140}
}
\authorrunning{S.~Y.~Pradata et al.}
%
\institute{
Department of Computer Science and Electronics, Faculty of Mathematics and Natural Sciences, Universitas Gadjah Mada, Yogyakarta 55281, Indonesia \email{muhammad.alfian.amrizal@ugm.ac.id} \and
Research Center for Quantum Physics, National Research and Innovation Agency (BRIN), South Tangerang 15314, Indonesia
\and
Department of Physics, Faculty of Mathematics and Natural Sciences, Universitas Gadjah Mada, Yogyakarta 55281, Indonesia
\and
Cyberscience Center, Tohoku University, Sendai 9800845, Japan
}
\maketitle              
\begin{abstract}
Classical simulations of quantum circuits are essential for verifying and benchmarking quantum algorithms, particularly for large circuits, where computational demands increase exponentially with the number of qubits.  Among available methods, the classical simulation of quantum circuits inspired by density functional theory---the so-called QC-DFT method, shows promise for large circuit simulations as it approximates the quantum circuits using single-qubit reduced density matrices to model multi-qubit systems.  However, the QC-DFT method performs very poorly when dealing with multi-qubit gates.  In this work, we introduce a novel CNOT ``functional" that leverages neural networks to generate unitary transformations, effectively mitigating the simulation errors observed in the original QC-DFT method.  For random circuit simulations, our modified QC-DFT enables efficient computation of single-qubit marginal measurement probabilities, or single-qubit probability (SQPs), and achieves lower SQP errors and higher fidelities than the original QC-DFT method.  Despite some limitations in capturing full entanglement and joint probability distributions, we find potential applications of SQPs in simulating Shor's and Grover's algorithms for specific solution classes.  These findings advance the capabilities of classical simulations for some quantum problems and provide insights into managing entanglement and gate errors in practical quantum computing.

\end{abstract}
%
%
%
\section{Introduction}

In the last few years, classical simulations of quantum circuits have emerged as a significant focus within the field of quantum computing~\cite{bravyi2016improved,jozsa2006simulation,jozsa2008matchgates,chen2018classical,kissinger2022classical,terhal2002classical,napp2022efficient,noh2020efficient,qassim2019clifford} because these simulations are invaluable for advancing scalable and reliable quantum technologies~\cite{jones2019quest}.  They serve multiple purposes, including the verification of quantum algorithms~\cite{broadbent2015verify}, the development of new quantum strategies~\cite{jones2019quest}, and the benchmarking of quantum advantages over classical computation~\cite{magesan2012characterizing}.  Additionally, classical simulations are essential for creating and refining error correction techniques and noise models, which are critical for mitigating the challenges posed by decoherence and operational errors in quantum systems~\cite{katsuda2024simulation}.  By providing a platform for researchers to explore complex quantum phenomena and assess the performance of quantum hardware, classical simulations play a pivotal role in bridging the gap between theoretical research and practical implementation of quantum computing in the future.

Although classical simulations are instrumental in advancing quantum computing, exact (classical) simulations of quantum circuits become infeasible as the number of qubits $N_q$ increases due to the well-known exponential $\mathcal{O}(2^{N_q})$ growth in computational and memory demands~\cite{jones2019quest}.  As each additional qubit doubles the state space that needs to be represented, simulating large quantum circuits quickly outstrips the capabilities of even the most powerful classical supercomputers.  Various methods have been developed in the last decade to address this challenge for efficient classical simulation techniques that can approximate quantum behavior for \emph{certain types} of problems or \emph{specific classes} of quantum circuits~\cite{bravyi2016improved,chen2018classical,kissinger2022classical,terhal2002classical,napp2022efficient,noh2020efficient}. These methods often leverage insights into the structure of particular quantum circuits or exploit symmetries within the system~\cite{kissinger2022classical,terhal2002classical,napp2022efficient,qassim2019clifford}, hopefully allowing researchers to simulate larger quantum systems than would otherwise be possible.

Recently, Bernardi~\cite{bernardi2023efficient} introduced a density functional theory (DFT)-inspired non-exact simulation of quantum circuits (QC); hereafter, we referred to this method as ``QC-DFT".  In particular, Bernardi proposed QC-DFT with a single-qubit reduced density matrix (1-RDM) approach that models the whole quantum system as a collection of single-qubit subsystems, represented by several 1-RDMs, and thus obtained constant time and linear space complexity with respect to the number of qubits.  The Bernardi's QC-DFT method is briefly outlined as follows.  Suppose we have a state vector at step $s$, i.e., $|\Psi_s\rangle$, from the circuit with $N_q$ qubits.  For each qubit $n$, the 1-RDM can be computed by partial tracing the other $N_q-1$ qubits from the full density matrix of $|\Psi_s\rangle$, i.e.,
\begin{equation}\label{1RDM_def}
    \rho^{(n)}_s = \text{Tr}_{j\neq n}(|\Psi_s\rangle\langle \Psi_s|),
\end{equation}
with $ j \in \{ 1, 2, \ldots, N_q\}$.
One can then define the single-qubit probability (SQP) of qubit $n$ at step $s$ as the marginal probability $p^{(n)}_s$ of measuring qubit $n$ in state $|1\rangle$ regardless of the state of the other qubits.  Given the state vector $|\Psi_s\rangle$, SQP can be calculated as
\begin{equation}
    p^{(n)}_s = \sum_{i_q,q\neq n}|\langle i_1,i_2,\hdots,i_n=1,\hdots,i_{N_q}|\Psi_s\rangle|^2 .
\end{equation}
Once we have the 1-RDM of the qubit $n$, the SQP is equal to the lower right diagonal component of the 1-RDM based on its definition in Eq.~(\ref{1RDM_def}).

In Bernardi's QC-DFT, one should start with a collection of 1-RDMs ($\rho^{(1)}_{0},\rho^{(2)}_{0},\hdots,\rho^{(N_q)}_{0}$) for each qubit and then evolve each 1-RDM separately.  Each qubit can be initialized at the state $|0\rangle$ corresponding to the 1-RDM:
\begin{equation}
    \rho^{(n)}_0 =
    \begin{pmatrix}
        1 & 0 \\
        0 & 0 \\
    \end{pmatrix}.
\end{equation}
Any single-qubit gate $\hat{U}^{(n)}_s$ applied on qubit $n$ at step $s$ can be \emph{exactly simulated} by
\begin{equation}
    \rho^{(n)}_{s+1} = \hat{U}^{(n)}_s \rho^{(n)}_{s} \hat{U}^{(n)\dag}_s .
\end{equation}
On the other hand, for the multi-qubit gates, specifically the CNOT gates, each is simulated by first taking the tensor product of the 1-RDM of the control and target qubit.  Then, we can apply the CNOT evolution matrix $U_{CX}$ and retake the partial trace of 1-RDM.  These steps are summarized as the so-called ``CNOT functional" (which, according to Bernardi~\cite{bernardi2023efficient}, is analogous to the exchange-correlation functional in DFT):
\begin{align}
    \rho_{s+1}^{(c)} &= \text{Tr}_t \left[ \hat{U}_{CX} \left( \rho_{s}^{(c)} \otimes \rho_{s}^{(t)} \right) \hat{U}_{CX}^\dag \right], \\
    \rho_{s+1}^{(t)} &= \text{Tr}_c \left[ \hat{U}_{CX} \left( \rho_{s}^{(c)} \otimes \rho_{s}^{(t)} \right) \hat{U}_{CX}^\dag \right],
\end{align}
where $\rho_i$ is the 1-RDM at step $i$ (after applying the $i$-th gate) and the superscripts $(c)$ and $(t)$ denote the control and target qubits, respectively.  \emph{However}, the 1-RDM of the control and target qubits resulting from this CNOT-functional does not agree with those obtained from the exact simulation, leading to simulation errors growing with the circuit depth.

\section{METHODS}

In this work, to significantly minimize simulation errors, we propose a modification of the CNOT functional by applying an intermediary unitary gate $\hat{U}_{m}$ before the CNOT gate as an attempt to correct the error.  This proposed CNOT functional can be written as
\begin{align}
    \rho_{s+1}^{(c)} &= \text{Tr}_t \left[ \hat{U}_{CX} {\hat{U}_{m}^{(c)}} \left( \rho_{s}^{(c)} \otimes \rho_{s}^{(t)} \right) {\hat{U}_{m}^{(c)\dag}} \hat{U}_{CX}^\dag \right], \\
    \rho_{s+1}^{(t)} &= \text{Tr}_c \left[ \hat{U}_{CX} {\hat{U}_{m}^{(t)}} \left( \rho_{s}^{(c)} \otimes \rho_{s}^{(t)} \right) {\hat{U}_{m}^{(t)\dag}} \hat{U}_{CX}^\dag \right] ,
\end{align}
where $\hat{U}_{m}$ is constructed by taking the imaginary exponent of a Hermitian matrix involving a set $P$ of Pauli matrices and parametrized by $\bm{\theta} = \left( \theta_{ij} \right)_{i,j}$ as follows,
\begin{align}
    \hat{U}_{m} &= e^{iH\left(\bm{\theta}\right)}, \\
    H\left(\bm{\theta}\right) &= \sum_{\sigma_i, \sigma_j \in P} \theta_{ij} \sigma_i \otimes \sigma_j, & P = \{ I, \sigma_x, \sigma_y, \sigma_z \}.
\end{align}
Note that the only information we can use to construct $\hat{U}_{m}$ is the 1-RDMs of the control and target qubits.  To overcome this problem, we also propose a neural network model to predict $\hat{U}_{m}$ based on the 1-RDMs.  Specifically, we want a pair of neural networks to approximate the functions 
\begin{align}
    f^{(c)} : \left( \rho_{s}^{(c)}, \rho_{s}^{(t)} \right) \mapsto \bm{\theta}^{(c)}, \\
    f^{(t)} : \left( \rho_{s}^{(c)}, \rho_{s}^{(t)} \right) \mapsto \bm{\theta}^{(t)},
\end{align}
so that each of the neural networks outputs a layer consisting of 16 neurons and each neuron corresponds to a value of $\theta_{ij}$.

The neural networks for predicting $\hat{U}_{m}$ can be trained with traditional backpropagation if provided with the appropriate loss function and its corresponding gradient.  In this case, since we want the predicted 1-RDMs ($\rho_{\mathrm{pred}}$) acquired through the proposed CNOT functional to be as close as possible to the exact 1-RDMs ($\rho_{\mathrm{exact}}$) acquired through exact quantum circuit simulations, we propose the \emph{infidelity} ($1 - F$) as a loss function, where the fidelity between two density matrices $\rho_1$ and $\rho_2$ is defined as~\cite{jozsa1994fidelity,nielsen2010quantum}
\begin{equation}
    F(\rho_1,\rho_2) = \text{Tr}\left[ \sqrt{\sqrt{\rho_1}\rho_2\sqrt{\rho_1}} \right].
\end{equation}
The fidelity ranges between 0 for orthogonal states and 1 for indistinguishable states so that the infidelity is suitable for the loss function. To average the infidelity over the training data, we use the root-mean-squared infidelity (RMS1F) error which we define as
\begin{equation}
    \mathcal{E}_{\mathrm{RMS1F}} = \sqrt{\frac{1}{N_d}\sum_{l=1}^{N_d} \left( 1 - F\left( \rho_{\mathrm{exact}}^{[l]}, \rho_{\mathrm{pred}}^{[l]} \right) \right)^2} ,
\end{equation}
where $N_d$ is the number of training data and $\rho^{[l]}$ is the $l$-th density matrix in the training data. The gradient of this loss function with respect to the unitary parameters $\bm{\theta}$ can be calculated analytically and is detailed in Appendix~\ref{app-a}.  Finally, we define two metrics to quantify the advantage of our proposed CNOT functional: (1) the SQP error and (2) the mean fidelity, given by
\begin{equation}
    \mathcal{E}_{\mathrm{SQP}} = \sqrt{\frac{1}{N_q} \sum_{k=1}^{N_q} \left( p^{(k)}_{\mathrm{exact}} - p^{(k)}_{\mathrm{pred}} \right)^2} \label{SQPE}
\end{equation}
and
\begin{equation}
    \bar{F} = \frac{1}{N_q} \sum_{k=1}^{N_q} F(\rho^{(k)}_{\mathrm{exact}},\rho^{(k)}_{\mathrm{pred}}), \label{MF}
\end{equation} 
respectively, where $p^{(k)}_{\textrm{exact}}$ is the SQP from the exact quantum circuit simulation while $p^{(k)}_{\textrm{pred}}$ is the SQP predicted from QC-DFT with the proposed CNOT functional. 

\section{Results and Discussion}

In the first step of training the neural network, we generate $N_d = 300$ mixed-state 2-qubit density matrices as the training data. We take the partial trace of these density matrices before and after applying the CNOT gate to get the exact 1-RDMs.  We then train the neural network by performing a simple and small search space hyperparameter tuning to find a sufficiently good model. In the numerical experiments, the best model found in our search space is trained for $500$ epochs with stochastic gradient descent optimizer~\cite{robbins1951stochastic} and learning rate on the order of $10^{-5}$.  The neural network consists of $8$ dense layers with $\left[ 5, 64, 64, 128, 256, 512, 1024, 16 \right]$ neurons and sigmoid activation functions. Readers interested in the results of the neural network training can check Appendix~\ref{app-b}.

\subsection{SQP and fidelity}

Based on the SQP error and mean fidelity of all qubits, we will show that the CNOT functional achieves a better approximation to the exact simulation than the original Bernardi CNOT functional.  As a testbed, we generate random quantum circuits with $N$ qubits, in which a single-qubit or two-qubit gate is applied to randomly chosen qubit(s) at each step.  The gates are composed of random gates sampled uniformly from the set $\{H, X, Y, Z, RX, RY, RZ, CNOT\}$---except for the CNOT gate, which has 8 times higher probability to be chosen over the other gates---with uniformly distributed parameters $r \in \left[ 0, 2\pi \right]$ for the rotation gates ($RX,RY,RZ$).  We run 300 different random circuits and then take the mean fidelity and SQP error average from all circuits at each step.    

\begin{figure}[H]
\centering\includegraphics[clip,width=8cm]{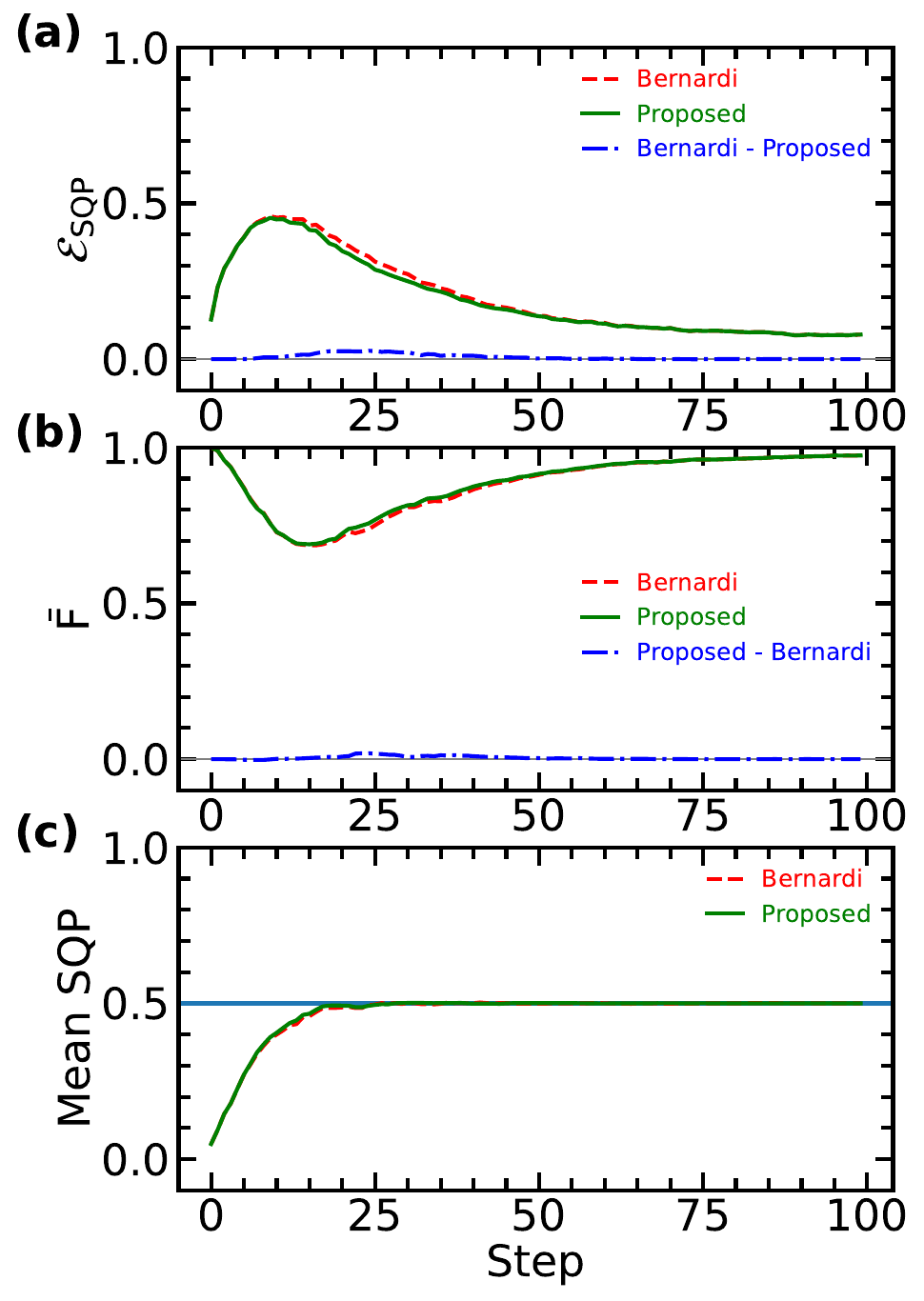}
\caption{\label{fig:SQP-Fidelity} Evolution of (a) SQP error, (b) mean fidelity, and (c) mean SQP calculated with our proposed CNOT functional (solid line) compared with the original QC-DFT method by Bernardi (dashed line).  In panels (a) and (b) the dash-dotted lines show the differences between the two methods, while gray lines are a guide for eyes to show zeros.  In panel (c), the horizontal solid line indicates the value of converged mean SQP at 0.5 (or 50\%).}
\end{figure}

Figures~\ref{fig:SQP-Fidelity}(a) and~\ref{fig:SQP-Fidelity}(b), respectively, show the average SQP error and fidelity calculated by our proposed method and Bernardi's QC-DFT in each step after applying the aforementioned random gates.  Our proposed method achieves better performance (lower SQP error) than Bernardi's QC-DFT at the steps where the blue dash-dotted line is above zero, as shown in Fig.~\ref{fig:SQP-Fidelity}(a).  Figure~\ref{fig:SQP-Fidelity}(b) also emphasizes that at each step, the mean fidelity of the 1-RDMs of our proposed method is consistently higher than that of Bernardi's QC-DFT.  Thus, it signifies the improvement posed by our proposed method.  Moreover, our proposed method executes at the same timescale as Bernardi's QC-DFT (Appendix~\ref{app-c}), indicating the speedup compared to the exact calculation.  Interestingly, the difference in SQP error converges to zero after many steps despite the SQP error being non-zero. This can be explained by the fact that the SQP of our proposed method converges to 50\%, as can be seen in Fig.~\ref{fig:SQP-Fidelity}(c). This phenomenon will be discussed as one of limitations of the proposed method.

\subsection{Limitations}

There are some limitations we found in the proposed method. Firstly, there is \emph{loss of entanglement information}.  Since we evolve the 1-RDM of each qubit separately, the entanglement information between different qubits cannot be captured.  The entanglement information cannot be retrieved even if we have a CNOT functional that accurately calculates the 1-RDM of the control and target qubits.  Using the 1-RDM, we can only determine the marginal measurement probability of a particular qubit collapsing into $|0\rangle$ or $|1\rangle$, regardless of the states of the other qubits.  Meanwhile, if that particular qubit is entangled with the other qubits, the probability depends on the measured state of those entangled qubits.  The entanglement information might be reassembled if we can reconstruct the full density matrix from the sequence of 1-RDMs of each qubit.  However, the full density matrix reconstruction is impossible since the mapping between a full density matrix to 1-RDMs is not bijective, i.e., some different density matrices may have the same set of 1-RDMs.  On the other hand, the mapping between unentangled full density matrices to 1-RDM sequences is bijective.  Therefore, the reconstruction of a full density matrix from 1-RDMs is only possible when the qubits are fully unentangled. In this case, it can be done by taking the tensor product of the 1-RDMs, $\rho = \sbigotimes_{n=1}^{N_q}\rho^{(n)}$.

The second limitation is \emph{the uncomputability of joint measurement probabilities}. Let the joint measurement probability of a quantum state collapsing to $\mathbf{x} = (x_1, x_2, \ldots, x_{N_q})$ be $P(\mathbf{x})$, and
\begin{align}
    \delta(x_i) = 
    \begin{cases} 
        1 - p^{(i)}, &x_i=0  \\ 
        p^{(i)}, &x_i=1 
    \end{cases}
\end{align}
On fully unentangled states, the SQP of each qubit $p^{(i)}$ is mutually independent. Hence, we can simply calculate the joint measurement probability of the system as $P(\mathbf{x}) = \prod_{i=1}^{N_q} \delta(x_i)$. However, entanglement introduces non-independence between the SQP of each qubit. When this happens, we can no longer construct the joint probability of the system.

The third limitation is the \emph{uncomputability of expectation values for entangled states}. Suppose we want to calculate the expectation value of an $N_q$-qubit observable ($\hat{O}$) constructed from the tensor product of single-qubit observables $\hat{O} = \sbigotimes_{i=1}^{N_q} \hat{O}^{(i)}$.  In exact quantum circuit simulations, given the full density matrix $\rho$, the expectation value can be calculated as $\langle \hat{O} \rangle = \text{Tr}\left[\rho\hat{O}\right]$. In our proposed method, we are only able to accurately compute $\langle \hat{O} \rangle$ if $\rho$ is a product state $\rho_{\otimes} = \sbigotimes_{i=1}^{N_q} \rho^{(i)}$.  In this case, 
\begin{align}
    \langle \hat{O} \rangle &= \text{Tr}\left[\rho_{\otimes} \hat{O}\right]
    \nonumber \\
    &= \text{Tr}\left[\sbigotimes_{i=1}^{N_q} \rho^{(i)}\hat{O}^{(i)}\right]
    = \prod_{i=1}^{N_q} \text{Tr}\left[\rho^{(i)}\hat{O}^{(i)}\right]
\end{align}
Note that the method fails if $\rho$ is an entangled state since we are unable to construct $\rho$ with just the 1-RDMs ($\rho^{(i)}$). This problem leads to the uncomputability of the expectation value of any entangled state.

Lastly, there is a problem of \emph{SQP convergence}. We observe that the SQP in the QC-DFT circuit tends to converge to 50\% after many steps. 
To explain this phenomenon, we prove that when applied to product states, the CNOT gate causes the SQP of the target qubit to approach 50\%. The full derivation can be seen in Appendix~\ref{app-d}.  We summarize the result as follows. First, the target SQP $p^{(t)}_s$ evolves to $p^{(t)}_{s+1}$ based on $p^{(t)}_s$ and the control SQP $p^{(c)}_s$ according to 
\begin{align}
    p^{(t)}_{s+1} &= p^{(t)}_s + p^{(c)}_s (1 - 2 p^{(t)}_s)
\end{align}
This equation resembles the linear interpolation of $p^{(t)}_{s+1}$ between $p^{(t)}_s$ to $1 - p^{(t)}_s$ based on the control SQP $p^{(c)}_s$. Since $\left| 0.5 - p^{(t)}_s\right| = \left| 0.5 - \left(1 - p^{(t)}_s \right) \right|$, $p^{(t)}_{s+1}$ never becomes further to $0.5$ than $p^{(t)}_s$.  Finally, since the SQP of the qubits could only get closer to 50\%, they will eventually converge to 50\% after the CNOT gates are applied multiple times.

\subsection{Potential applications}

Despite the limitations, we find that there are some possibly significant use cases of SQPs.  In particular, specific problems can still be answered without the joint measurement probability of the qubits.  The solutions to such problems are usually encoded as bitstrings resulting from measurements of quantum states at the end of the circuits.  Therefore, once the SQP convergence problem is resolved, our method will be able to efficiently simulate certain classes of quantum algorithms.

\subsubsection{Grover's algorithm}

As the first application, in Grover's algorithm, the circuit amplifies the amplitude of the solution state. To demonstrate this, we show the exact SQP results on Grover's algorithm for the first $5$ iterations of applying the oracle and diffusion operator with the solution $10110$ in Table.~\ref{Table:Grover-Single-Solution}. The SQP of the qubits whose solution is $1$ increases, while the SQP of the qubits whose solution is $0$ decreases. Hence, if there is only a single solution to the search, one can simply determine the solution by rounding the SQP of each qubit. In fact, in classical simulations, this rounding method can be done after only one application of the oracle and diffusion operator in the circuit, minimizing the need to repeat the oracle and diffusion operator.

\begin{table}[!b]
\centering
\caption{The SQP values for each iteration of Grover's algorithm with a single solution $10110$. Qubit 4 is the most significant bit.}
\label{Table:Grover-Single-Solution} 
\begin{tabular}{|c||c|c|c|c|c|}
\hline
 & \multicolumn{5}{c|}{\textbf{Iteration}} \\
\hline
\textbf{Qubit} & 0 & 1 & 2 & 3 & 4 \\
\hline
4 & 0.5 & 0.6171875 & 0.7947998 & 0.94680595 & 0.999577969 \\
3 & 0.5 & 0.3828125 & 0.2052002 & 0.05319405 & 0.000422031 \\
2 & 0.5 & 0.6171875 & 0.7947998 & 0.94680595 & 0.999577969 \\
1 & 0.5 & 0.6171875 & 0.7947998 & 0.94680595 & 0.999577969 \\
0 & 0.5 & 0.3828125 & 0.2052002 & 0.05319405 & 0.000422031 \\
\hline
\end{tabular}
\end{table}

In the case of multi-solution Grover's search, SQPs alone are generally insufficient to identify all solutions. However, when there are only two solutions that differ by a single bit, SQP can be used to find them. In Table~\ref{Table:Grover-Multi-Solution}, we present an example of the SQPs in Grover's search for solutions $10110$ and $10111$. The SQPs for qubits with identical solutions in both cases behave similarly to the single-solution scenario. Meanwhile, the SQP for the one qubit that differs between the two solutions will remain at $0.5$ (or $50\%$).

\begin{table}[!t]
\centering
\caption{The SQP values for each iteration of Grover's algorithm with multiple solutions $10110$ and $10111$.  The SQP can be used here because the solutions only differ by a single bit. On the last iteration, over-rotation happens and causes the SQP to stray from the solution.}
\label{Table:Grover-Multi-Solution} 
\begin{tabular}{|c||c|c|c|c|c|}
\hline
 & \multicolumn{5}{c|}{\textbf{Iteration}} \\
\hline
\textbf{Qubit} & 0 & 1 & 2 & 3 & 4   \\
\hline
4 & 0.5 & 0.71875 & 0.95117187 & 0.97937012 & 0.77690887 \\
3 & 0.5 & 0.28125 & 0.04882812 & 0.02062988 & 0.22309113 \\
2 & 0.5 & 0.71875 & 0.95117187 & 0.97937012 & 0.77690887 \\
1 & 0.5 & 0.71875 & 0.95117187 & 0.97937012 & 0.77690887 \\
0 & 0.5 & 0.5 & 0.5 & 0.5 & 0.5 \\
\hline
\end{tabular}
\end{table}

\subsubsection{Shor's algorithm}

We can also apply the SQPs for Shor's algorithm.  It is known that Shor's algorithm allows us to factorize a semiprime $N_s$ into two prime numbers.  The quantum component of the algorithm is used to determine $r$, i.e., the period of a given integer $a \mod N_s$.  This period $r$ can then be leveraged to find the factors of $N_s$ through some classical post-processing.  Notably, when $r = 2^y$ for a positive integer $y$, it can be shown that measuring the first register in the quantum subroutine results in a state $\ket{k}$, where $k = 2^{2n_s - y}$ and $n_s$ is the number of qubits required to satisfy $2^{n_s} \geq N_s$~\cite{shor1994algorithms,mermin2007quantum,vathsan2015introduction,nielsen2010quantum}. In this scenario, the first $2n_s - y_0$ qubits will have zero SQP (or close to zero due to numerical precision errors).  Thus, $y_0$ can be determined by counting the qubits with non-zero SQP, allowing us to calculate $r = 2^{y_0}$. If we can show that there is at least one value of $a$ such that $r = 2^y$ for any arbitrary semiprime $N_s$, then we can factorize $N_s$ with only SQPs (without the joint probability of the quantum system).

\begin{figure}[H]
\centering\includegraphics[clip,width=6cm]{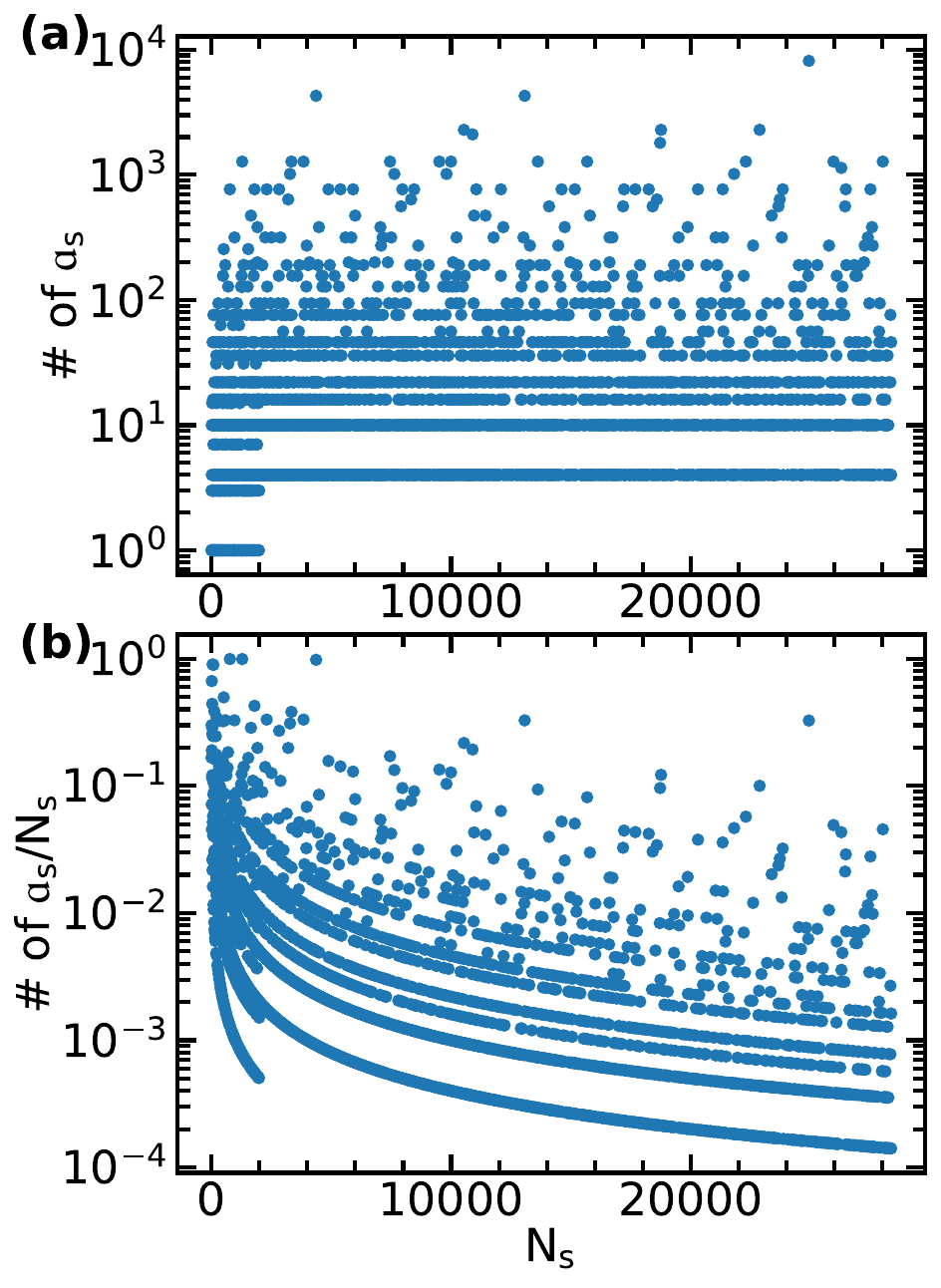}
\caption{\label{fig:Shor} The success rate of simulating Shor's algorithm using SQPs. (a) The number of integers $1 \leq a_s \leq n_S$ for which the period of $a_s \mod N_s$ satisfies $r = 2^y$ and $\gcd\left(a_s^{r/2} + 1, N_s\right)$ yields a non-trivial factor for some $N_s$. (b) The probability of sampling a value $a_s$ from arbitrary integers $a \in \left[1, N_s\right]$ for the first 3000 squarefree semiprimes $N_s$.}
\end{figure}

The values of $a$ for which (1) the period of $a \mod N_s$ satisfies $r = 2^y$ and (2) $\gcd\left(a^{r/2} + 1, N_s\right)$ yields a non-trivial factor for some $N_s$ will be denoted as $a_s$.  We conducted a brute-force calculation for the number of $a_s$ values for the first 100 square semiprimes.  For all of these square semiprimes, we found no valid values for $a_s$. Thus, when attempting to factorize $N_s$ using SQP, it is essential to verify that $N_s$ is not a square semiprime. This check can be performed by determining whether $\sqrt{N_s}$ is an integer.   Furthermore, we conducted a similar brute-force calculation for the first 3,000 squarefree semiprimes.  As shown in Fig.~\ref{fig:Shor}(a), even at lower values of $N_s$, there is at least one $a_s$ value.  Groups of semiprimes tend to have consistent numbers of $a_s$ values, indicated by horizontal line patterns in the figure.  Additionally, the count of $a_s$ values shows a very slight increase as $N_s$ grows.

Figure~\ref{fig:Shor}(b) illustrates the probability $p_r(a_s)$ of obtaining $a_s$ by randomly sampling an integer $a \in \left[1, N_s\right]$, where $p_r(a_s) = \text{count}(a_s)/N_s$. We observe that $p_r(a_s)$ reciprocally decays as $N_s$ increases, suggesting that finding $a_s$ values becomes more challenging for larger $N_s$. Despite this trend, it remains feasible to use SQPs to factorize squarefree semiprimes.  Complemented by the fact that square semiprimes can be easily factorized by taking the square root, this observation implies that it is possible to factorize most semiprimes with just SQPs. However, while we have numerically demonstrated that at least one $a_s$ value exists for each of the first 3000 semiprimes, we have not rigorously proven that such values exist for any arbitrary squarefree semiprime, which can be an open problem for the future.

\section{Conclusions}

We have developed a significantly improved QC-DFT method with a new CNOT functional which gives an extra error-correcting intermediary unitary gate generated by a neural network based on the input 1-RDMs.  With this approach, we have achieved lower SQP error and higher fidelity compared to the original QC-DFT.  Despite the improvements, there are still inherent limitations to our method, especially the loss of entanglement information that leads to the inability to calculate expectation values and construct the joint probability of the system.  Due to the nature of the CNOT functional, the SQPs also tend to converge to 50\% after applying multiple CNOT gates.  Nevertheless, we succeeded in exploring a handful of possible utilization of the SQPs, particularly on the specific classes of solutions of Grover's and Shor's algorithms.

\begin{credits}
\subsubsection{\ackname} This work is partly supported by  JSPS
KAKENHI Grant Number JP24K02945. We thank Dr. Agung Budiyono (BRIN Research Center for Quantum Physics) for the fruitful discussion that led us to look for potential use cases of SQPs.  We acknowledge the Quasi Lab and Mahameru BRIN for their mini-cluster and HPC facilities.  SYP acknowledges Beasiswa Unggulan while MANA acknowledges Beasiswa Prestasi Talenta, both from the Ministry of Education, Culture, Research and Technology, Republic of Indonesia. HML is supported by the research assistantship from the National Talent Management System at BRIN.  Codes to reproduce all the results presented in this paper are available at \url{https://github.com/BRIN-Q/QCDFT-ML}.

\subsubsection{\discintname}
The authors have no competing interests to declare that are
relevant to the content of this article.
\end{credits}

\appendix
\section*{Appendix}
\vspace{-3pt}
\section{Gradient of The Loss Function}
\label{app-a}
We use infidelity ($1-F$) as the loss function for developing a neural network model that can give accurate CNOT functionals.  The gradient of the loss function with respect to the unitary parameters $\bm{\theta}$ can be calculated as shown below:
\begin{align}
    \frac{\partial \mathcal{E}_{\mathrm{RMS1F}}}{\partial \theta_{ij}} &= \frac{\partial}{\partial \theta_{ij}} \sqrt{\frac{1}{N_d}\sum_{l=1}^{N_d} \left( 1 - F\left( \rho^{[l]}_{\mathrm{exact}}, \rho^{[l]}_{\mathrm{pred}} \right) \right)^2} \nonumber\\
    &= \frac{1}{N_d \times \mathcal{E}_{\mathrm{RMS1F}}} \sum_{l=1}^{N_d} \left( F\left( \rho^{[l]}_{\mathrm{exact}}, \rho^{[l]}_{\mathrm{pred}}\right) - 1 \right) \frac{\partial}{\partial \theta_{ij}} F\left( \rho^{[l]}_{\mathrm{exact}}, \rho^{[l]}_{\mathrm{pred}} \right).
\end{align}
The term $\displaystyle\frac{\partial}{\partial \theta_{ij}} F\left( \rho^{[l]}_{\mathrm{exact}}, \rho^{[l]}_{\mathrm{pred}} \right)$ in Eq.~(S1) can be calculated as follows:
\begin{align}
    \frac{\partial}{\partial \theta_{ij}} F\left( \rho^{[l]}_{\mathrm{exact}}, \rho^{[l]}_{\mathrm{pred}} \right) &= \frac{\partial}{\partial \theta_{ij}} \text{Tr} \left[ \sqrt{\sqrt{\rho^{[l]}_{\mathrm{exact}}} \rho^{[l]}_{\mathrm{pred}} \sqrt{\rho^{[l]}_{\mathrm{exact}}}} \right] \nonumber\\
    \nonumber\\
    &= \frac{1}{2} \text{Tr} \left[ \sqrt{\sqrt{\rho^{[l]}_{\mathrm{exact}}} \rho^{[l]}_{\mathrm{pred}} \sqrt{\rho^{[l]}_{\mathrm{exact}}}}^{-1} \sqrt{\rho^{[l]}_{\mathrm{exact}}} \left( \frac{\partial}{\partial \theta_{ij}} \rho^{[l]}_{\mathrm{pred}} \right) \sqrt{\rho^{[l]}_{\mathrm{exact}}} \right].
\end{align}
Finally, the term $\displaystyle\frac{\partial}{\partial \theta_{ij}} \rho^{[l]}_{\mathrm{pred}}$ can be calculated as follows:
\begin{align}
    \frac{\partial}{\partial \theta_{ij}} \rho^{[l]}_{\mathrm{pred}} =&  
    \, \frac{\partial}{\partial \theta_{ij}} \text{Tr}_{c/t} \left[ U_{CX} U_m^{(t)/(c)} \left( \rho_{s}^{(c)} \otimes \rho_{s}^{(t)} \right) U_m^{(t)/(c)\dag} U_{CX}^\dag \right] \nonumber\\
    =& \, \text{Tr}_{c/t} \left[ U_{CX} \left( \frac{\partial}{\partial \theta_{ij}}  e^{iH\left(\bm{\theta}\right)} \right) \left( \rho_{s}^{(c)} \otimes \rho_{s}^{(t)} \right) e^{-iH^\dag\left(\bm{\theta}\right)} U_{CX}^\dag \right] \nonumber\\
    &+\text{Tr}_{c/t} \left[ U_{CX} e^{iH\left(\bm{\theta}\right)} \left( \rho_{s}^{(c)} \otimes \rho_{s}^{(t)} \right) \left(\frac{\partial}{\partial \theta_{ij}} e^{-iH^\dag\left(\bm{\theta}\right)} \right) U_{CX}^\dag \right],
\end{align}
where
\begin{align}
    \frac{\partial}{\partial \theta_{ij}}e^{iH\left(\bm{\theta}\right)} &= e^{iH\left(\bm{\theta}\right)} \frac{\partial}{\partial \theta_{ij}} iH\left(\bm{\theta}\right) \nonumber\\
    &= i U_m \frac{\partial}{\partial \theta_{ij}} \sum_{\sigma_i, \sigma_j \in P} \theta_{ij} \sigma_i \otimes \sigma_j, \text{~~~~~~~} P = \{ I, \sigma_x, \sigma_y, \sigma_z \} \nonumber \\
    &= i U_m \sigma_i \otimes \sigma_j,
\end{align}
and
\begin{align}    
    \frac{\partial}{\partial \theta_{ij}} e^{-iH^\dag\left(\bm{\theta}\right)} &= \frac{\partial}{\partial \theta_{ij}} e^{-iH\left(\bm{\theta}\right)} = - i U_m \sigma_i \otimes \sigma_j.
\end{align}


\section{Neural Network Training Results}
\label{app-b}

\begin{figure}[b]
\centering\includegraphics[clip,width=0.95\linewidth]{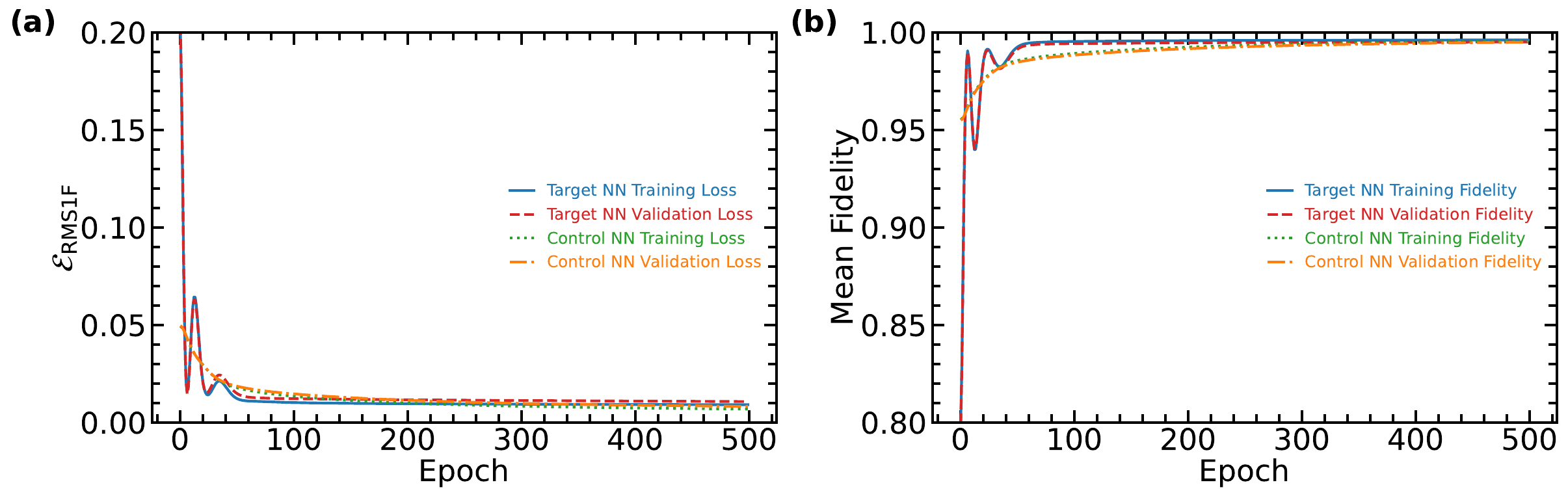}
\caption{\label{fig:Training} The evolution of (a) the loss function and (b) the mean fidelity during the training of the neural networks.}
\end{figure}

We train our neural network by performing a simple and small search space hyperparameter tuning to find a sufficiently good model.  In the numerical experiments as shown in Fig.~\ref{fig:Training} for the loss function and the mean fidelity, the best model found in our search space is trained for $500$ epochs with stochastic gradient descent optimizer~\cite{robbins1951stochastic} and learning rate on the order of $10^{-5}$.  The neural network consists of $8$ dense layers with $\left[ 5, 64, 64, 128, 256, 512, 1024, 16 \right]$ neurons and sigmoid activation functions.


\section{Time Benchmark}
\label{app-c}

\begin{figure}[h!]
\centering\includegraphics[clip,width=0.65\linewidth]{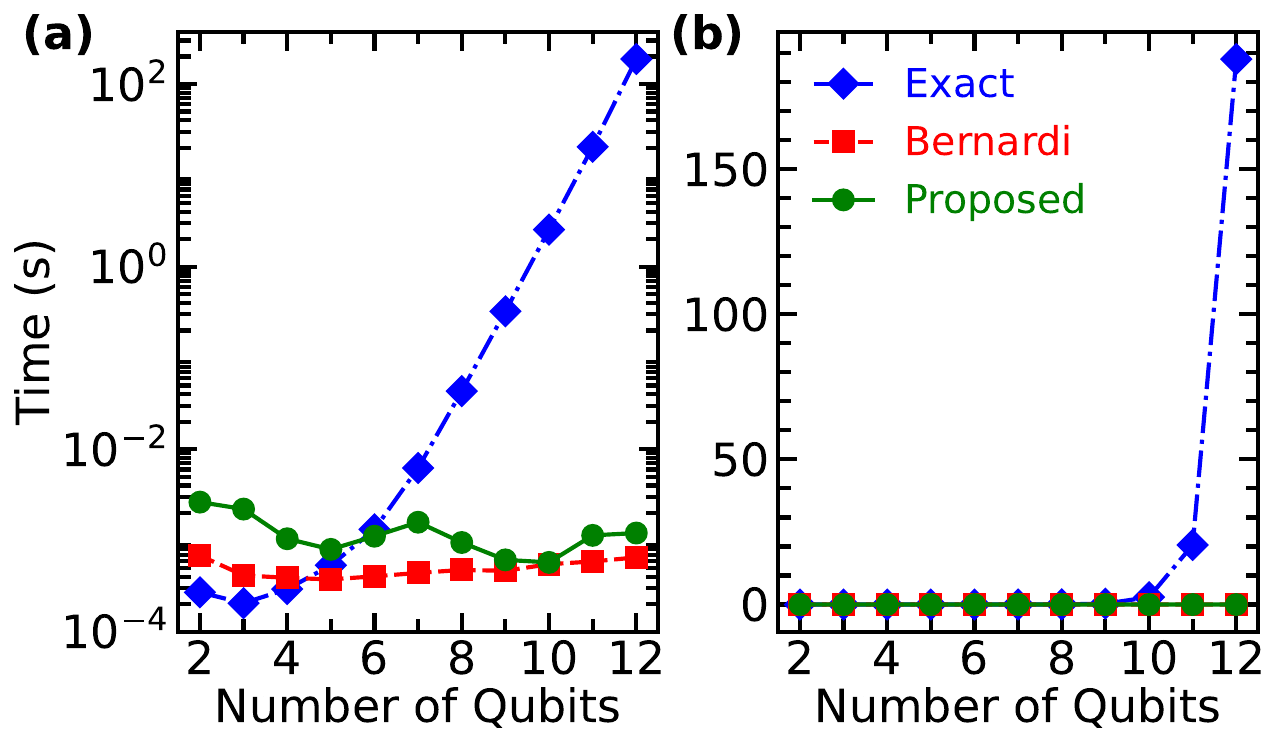}
\caption{\label{fig:Benchmark} Time Benchmark for the three different methods. Each data point is the average time it takes to apply 10 random gates for 10 circuits. }
\end{figure}

When comparing the calculation time between our proposed method with the exact calculation and the original Bernardi's QC-DFT, we found that our method can execute at the same timescale as Bernardi’s QC-DFT as shown in Fig.~\ref{fig:Benchmark}, indicating the speedup compared to the exact calculation. 

\section{Why CNOTs Cause SQPs to Converge to 50\%}
\label{app-d}
Let $\rho^{(c)}$ and $\rho^{(t)}$ be 1-RDMs. Let $p^{(c)}$ and $p^{(t)}$ be the Single Qubit Probability (SQP) for measuring $\ket{1}$. Since we are only interested in the measurement probabilities, we only need to take a look at the diagonal of the 1-RDMS.

\begin{align}
\begin{split}
    \rho^{(c)} &= 
    \begin{bmatrix}
        1 - p^{(c)} & - \\
        - & p^{(c)}
    \end{bmatrix}, \;\;\;\;\;\;
    \rho^{(t)} = 
    \begin{bmatrix}
        1 - p^{(t)} & - \\
        - & p^{(t)}
    \end{bmatrix} \\
\end{split}
\end{align}

\subsection{The effect of CNOT on the SQP of $\rho^{(c)}$ and $\rho^{(t)}$}
In this subsection, we attempt to obtain an expression for the SQP of the control and target qubits after the CNOT gate as the original SQP. First, we construct the tensor product of the 1-RDMs of both qubits, and apply the CNOT gate:
\begin{align}
    \rho^{(c)} \otimes \rho^{(t)} &= 
    \begin{bmatrix}
        \left(1 - p^{(c)}\right) \left(1 - p^{(t)}\right) & - & - & - \\
        - & \left(1 - p^{(c)}\right) p^{(t)} & - & - \\
        - & - & p^{(c)} \left(1 - p^{(t)}\right) & - \\
        - & - & - & p^{(c)} p^{(t)}
    \end{bmatrix} \nonumber
\end{align}
\begin{align}
    CNOT \left( \rho^{(c)} \otimes \rho^{(t)} \right) CNOT^\dag &=
    \begin{bmatrix}
        \left(1 - p^{(c)}\right) \left(1 - p^{(t)}\right) & - & - & - \\
        - & \left(1 - p^{(c)}\right) p^{(t)} & - & - \\
        - & - &  p^{(c)} p^{(t)} & - \\
        - & - & - & p^{(c)} \left(1 - p^{(t)}\right)
    \end{bmatrix} 
\end{align}
Next, we construct $\rho'^{(c)}$ and $\rho'^{(t)}$ back using partial trace:
\begin{align}
    \rho'^{(c)} &= \text{Tr}_t 
    \begin{bmatrix}
        \left(1 - p^{(c)}\right) \left(1 - p^{(t)}\right) & - & - & - \\
        - & \left(1 - p^{(c)}\right) p^{(t)} & - & - \\
        - & - &  p^{(c)} p^{(t)} &  \\
        - & - & - & p^{(c) \left(1 - p^{(t)}\right)}
    \end{bmatrix} \nonumber\\
    &= 
    \begin{bmatrix}
        1 - p^{(c)} & - \\
        - & p^{(c)}
    \end{bmatrix}
\end{align}

CNOT does not affect the SQP of $\rho^{(c)}$.

\begin{align}
    \rho'^{(t)} &= \text{Tr}_c 
    \begin{bmatrix}
        \left(1 - p^{(c)}\right) \left(1 - p^{(t)}\right) & - & - & - \\
        - & \left(1 - p^{(c)}\right) p^{(t)} & - & - \\
        - & - &  p^{(c)} p^{(t)} &  \\
        - & - & - & p^{(c)} \left(1 - p^{(t)}\right)
    \end{bmatrix} \nonumber \\
    &= 
    \begin{bmatrix}
        1 - p^{(c)} - p^{(t)} + 2 p^{(c)} p^{(t)} & - \\
        - & p^{(t)} + p^{(c)} - 2 p^{(c)} p^{(t)}
    \end{bmatrix} 
\end{align}

However, CNOT does affect the SQP of $\rho^{(t)}$.

\subsection{Upper and Lower Bounds of $\rho^{(t)}$ SQP}

Next, we want to show that the CNOT gate causes the SQPs to converge to $0.5$.  This can be done by proving that $\left| 0.5 - p^{(t)} \right| \geq \left| 0.5 - \left(p^{(t)} + p^{(c)} - 2 p^{(c)} p^{(t)}\right) \right|, \forall p^{(c)}, p^{(t)} \in \left[ 0, 1 \right]$, i.e., the SQP of pre-CNOT $\rho^{(t)}$ is strictly farther (or equally far) from 0.5 than the SQP of post-CNOT $\rho^{(t)}$, for all possible SQP values of pre-CNOT $\rho^{(t)}$ and $\rho^{(c)}$.  To prove that this equation holds, first, we rewrite the r.h.s. as follows:
\begin{align}
    \left(p^{(t)} + p^{(c)} - 2 p^{(c)} p^{(t)}\right) &= \left( -2 p^{(c)} + 1 \right) \left( p^{(t)} - 0.5 \right) + 0.5 \nonumber \\
    0.5 - \left(p^{(t)} + p^{(c)} - 2 p^{(c)} p^{(t)}\right) &= \left( 2 p^{(c)} - 1 \right) \left( p^{(t)} - 0.5 \right)
\end{align}
So, the inequality to prove now is
\begin{align}
    \left| 0.5 - p^{(t)} \right| \geq \left| \left( 2 p^{(c)} - 1 \right) \left( p^{(t)} - 0.5 \right) \right|, \forall p^{(c)}, p^{(t)} \in \left[ 0, 1 \right]
\end{align}

Next, we divide the inequality into several cases:

\begin{itemize}

    
    \item Case 1: $p^{(c)} \leq 0.5, p^{(t)} < 0.5$

    Because $0.5 - p^{(t)} \geq 0$ and $\left( 2 p^{(c)} - 1 \right) \left( p^{(t)} - 0.5 \right) \geq 0$
    , the inequality is equivalent to:
    \begin{align}
    \begin{split}
        \left| 0.5 - p^{(t)} \right| &\geq \left| \left( 2 p^{(c)} - 1 \right) \left( p^{(t)} - 0.5 \right) \right| \\
        0 &\leq p^{(c)} 
    \end{split}
    \end{align}
    The above inequality ($0 \leq p^{(c)}$) holds true $\forall p^{(c)}, p^{(t)} \in \left[ 0, 1 \right]$ with the case constraint $p^{(c)} \leq 0.5, p^{(t)} < 0.5$

    \item Case 2: $p^{(c)} \geq 0.5, p^{(t)} < 0.5$

    Because $0.5 - p^{(t)} \geq 0$ and $\left( 2 p^{(c)} - 1 \right) \left( p^{(t)} - 0.5 \right) \leq 0$
    , the inequality is equivalent to:
    \begin{align}
    \begin{split}
        \left| 0.5 - p^{(t)} \right| &\geq \left| \left( 2 p^{(c)} - 1 \right) \left( p^{(t)} - 0.5 \right) \right| \\
        1 &\geq p^{(c)} 
    \end{split}
    \end{align}
    The above inequality ($1 \geq p^{(c)}$) holds true $\forall p^{(c)}, p^{(t)} \in \left[ 0, 1 \right]$ with the case constraint $p^{(c)} \geq 0.5, p^{(t)} < 0.5$

    \item Case 3: $p^{(t)} = 0.5$
    The inequality is equivalent to:
    \begin{align}
    \begin{split}
        \left| 0.5 - p^{(t)} \right| &\geq \left| \left( 2 p^{(c)} - 1 \right) \left( p^{(t)} - 0.5 \right) \right| \\
        0 &\geq 0 
    \end{split}
    \end{align}
    The above inequality ($0 \geq 0$) holds true $\forall p^{(c)}, p^{(t)} \in \left[ 0, 1 \right]$ with the case constraint $p^{(t)} = 0.5$

    \item Case 4: $p^{(c)} \leq 0.5, p^{(t)} > 0.5$

    Because $0.5 - p^{(t)} \leq 0$ and $\left( 2 p^{(c)} - 1 \right) \left( p^{(t)} - 0.5 \right) \leq 0$
    , the inequality is equivalent to:
    \begin{align}
    \begin{split}
        \left| 0.5 - p^{(t)} \right| &\geq \left| \left( 2 p^{(c)} - 1 \right) \left( p^{(t)} - 0.5 \right) \right| \\
        p^{(c)} &\geq 0
    \end{split}
    \end{align}
    The above inequality ($p^{(c)} \geq 0$) holds true $\forall p^{(c)}, p^{(t)} \in \left[ 0, 1 \right]$ with the case constraint $p^{(c)} \leq 0.5, p^{(t)} > 0.5$

    \item Case 5: $p^{(c)} \geq 0.5, p^{(t)} > 0.5$

    Because $0.5 - p^{(t)} \leq 0$ and $\left( 2 p^{(c)} - 1 \right) \left( p^{(t)} - 0.5 \right) \geq 0$
    , the inequality is equivalent to:
    \begin{align}
    \begin{split}
        \left| 0.5 - p^{(t)} \right| &\geq \left| \left( 2 p^{(c)} - 1 \right) \left( p^{(t)} - 0.5 \right) \right| \\
        1 &\geq p^{(c)} 
    \end{split}
    \end{align}
    The above inequality ($1 \geq p^{(c)}$) holds true $\forall p^{(c)}, p^{(t)} \in \left[ 0, 1 \right]$ with the case constraint $p^{(c)} \geq 0.5, p^{(t)} > 0.5$
\end{itemize}

It is proven that the inequality holds true for all possible cases. Hence, the CNOT gate strictly causes the SQP to get closer to $0.5$. This means that if the CNOT gate is applied repeatedly, the SQPs will converge to $0.5$.

%
%
%
\bibliographystyle{splncs04}
\bibliography{references}
\end{document}